\documentclass[prd,aps,twocolumn,preprintnumbers, showpacs, nofootinbib,superscriptaddress,notitlepage]{revtex4-1}
\usepackage{amssymb,amsthm,amsmath}
\usepackage{textcomp}
\usepackage{subfig,graphicx}   
\usepackage{color}      
\usepackage{slashed}    
\usepackage{verbatim}
\usepackage[normalem]{ulem}
\usepackage{rotating}   
\usepackage{multirow}   
\begin{document}
\title{
Analysis of Charmed Baryon Decays with Strong Phases under SU(3) Symmetry
}

\affiliation{ Department of Physics and Institute of Theoretical Physics, Nanjing Normal University, Nanjing, Jiangsu 210023, China}
\affiliation{Particle Theory and Cosmology Group, Center for Theoretical Physics of the Universe, Institute for Basic Science (IBS), Daejeon 34126, Korea }
\author{Jin Sun$^{2}$}
\email{sunjin0810@ibs.re.kr}
\author{Zhi-Peng  Xing$^{1}$}
\email{zpxing@nnu.edu.cn(corresponding author)}  
\author{Ruilin Zhu$^{1}$}
\email{rlzhu@njnu.edu.cn}

\preprint{CTPU-PTC-24-18}

\begin{abstract}

The SU(3) analysis is considered a powerful tool in charmed baryon decays.
Motivated by recent measurements of anti-triplet charmed baryon two-body decays from the Belle, Belle II, and BESIII Collaborations, we have finally determined the last two IRA form factors, $f^{a\prime}$ and $g^{a\prime}$, which were absent in previous work.
By considering both real and complex form factor cases in our work, we find that the phases of the form factors are necessary and that complex form factors can explain the experimental data well.
Using the fitted form factors, we further numerically study the equivalence of the SU(3) 
irreducible representation amplitude (IRA) 
and topological diagrammatic approach (TDA) methods.
We find that the IRA and TDA methods can be numerically equivalent with the addition of the new form factors.
Based on the conclusions above, and considering some interesting scenarios, the CP violation effects can be predicted in these processes at the order of $O(10^{-3})$.
This suggests a promising opportunity to observe CP violation for the first time in charmed baryon decays.
Although our predictions depend on some assumptions, considering that the experimental data is far from sufficient to determine CPV through SU(3) analysis, our study is meaningful and instructive for observing CPV at experimental facilities.

\end{abstract}

\maketitle

\section{Introduction}

Over the past 20 years, with an increasing number of measurements by the BESIII, Belle, Belle II, and LHCb collaborations~\cite{Belle:2004zjl,Belle:2008xmh,LHCb:2013xam,
LHCb:2015tgy,BESIII:2015ysy,BESIII:2015bjk,LHCb:2017uwr,BESIII:2017kqg,Belle:2017ext,LHCb:2018pcs,LHCb:2018nfa}, charmed baryon decays, as an important part of heavy flavor physics, have attracted increasing attention from both the experimental and theoretical communities~\cite{Roberts:2007ni,Briceno:2012wt,Romanets:2012hm,Brown:2014ena,Lu:2016ogy,He:2018php,He:2018joe,Zhao:2018mov,He:2019tik,Geng:2019xbo,Jia:2019zxi}.
Due to the low production threshold, charmed baryons, including singly and doubly charmed baryon decays, have accumulated a significant amount of experimental data.
Meanwhile, the rich experimental phenomena in charmed baryon decays face difficulties in perturbative studies because the energy scale is far from the perturbative region.
However, the complicated non-perturbative effects also provide an opportunity to study the non-perturbative properties of quantum chromodynamics (QCD).
Therefore, charmed baryon decays are a promising platform for studying QCD and precisely testing the Standard Model (SM).

Among the various studies of charmed baryon decays, the SU(3) flavor symmetry analysis,as an non-perturbative method, including the irreducible representation amplitude (IRA) and the topological diagrammatic approach (TDA), is one of the most powerful methods for charmed baryon two-body decays~\cite{Lu:2016ogy,Zhao:2018mov,
Geng:2019xbo,Jia:2019zxi,Wang:2020gmn,Huang:2021aqu,Xing:2023dni,Geng:2023pkr,Zhong:2024qqs,Wang:2024ztg,Cheng:2024lsn}.
Without detailed dynamical understanding, predictions can be made based on the derived SU(3) relations or global analyses with sufficient experimental data.
The quality of the symmetry can be evaluated by how well the known data are explained under the imposed symmetry, as indicated by a good fit quality $\chi^2/\text{d.o.f}$.
As the simplest type of charmed baryon decays, the anti-triplet charmed baryon two-body decays have accumulated a large amount of experimental data and are therefore the most suitable processes for SU(3) analysis.
In previous studies, global fits show that SU(3) flavor symmetry in anti-triplet charmed baryon two-body decays is a good approximation~\cite{Xing:2023dni}.
However, due to the lack of key experimental data, the SU(3) parameters still cannot be fully determined~\cite{Xing:2023dni}.
Therefore, we look forward to additional measurements of key experimental data, which will enable us to conduct deeper and more comprehensive studies.

Fortunately, in 2024, the BESIII Collaboration measured the decay branching ratios of $\Lambda^{+}_{c}$, while the Belle and Belle II Collaborations measured the decay branching ratios and asymmetry parameters of $\Xi_c^0$ for the first time~\cite{BESIII:2023uvs,Belle-II:2024jql,BESIII:2024sfz}, as well as the decay branching ratios of $\Xi^{+}_{c}$~\cite{Belle-II:2024vax} as
\begin{eqnarray}
&&Br(\Lambda^{+}_{c}\to p \pi^0 ) =(0.0156^{+0.0072}_{-0.0058}\pm0.002)\%,\notag\\
&&Br(\Lambda^{+}_{c}\to p \eta ) =(0.163\pm0.031\pm0.011)\%,\notag\\
&&Br(\Lambda^{+}_{c}\to p K_L ) =(1.67\pm0.06\pm0.04)\%,\notag\\
&&Br(\Xi_c^0\to\Xi^0\pi^0)=(0.69\pm0.03\pm0.05\pm 0.13)\%,\nonumber\\
&&Br(\Xi_c^0\to\Xi^0\eta)=(0.16\pm0.02\pm 0.02\pm 0.03)\%,\nonumber\\
&&Br(\Xi_c^0\to\Xi^0\eta^{\prime})=(0.12\pm0.03\pm 0.01\pm 0.02)\%,\nonumber\\
&&Br(\Xi_c^+\to p K_S^0)=(7.16\pm0.46\pm0.20\pm3.21)\times10^{-4},\nonumber\\
&&Br(\Xi_c^+\to\Lambda\pi^+)=(4.52\pm0.41\pm0.26\pm2.03)\times10^{-4},\nonumber\\
&&Br(\Xi_c^+\to\Sigma^0\pi^+)=(1.20\pm0.08\pm0.07\pm0.54)\times10^{-3},\nonumber\\
&&\alpha(\Xi_c^0\to\Xi^0\pi^0)=-0.90\pm0.15\pm0.23.
\label{belle}
\end{eqnarray}
Based on the new measurements, all SU(3) IRA amplitudes can now be determined, enabling a deeper understanding of the weak interactions in charmed particle decays, including potential CP violation (CPV) effects.

CP violation (CPV) is a fundamental topic in particle physics~\cite{Lee:1957qs,Kobayashi:1973fv,Deshpande:1994ii,LHCb:2019hro,Lenz:2020awd,Schacht:2021jaz,Bediaga:2022sxw,Wang:2022tcm,Wang:2022fih,Maccolini:2023luv,Shen:2023eln,Song:2024jjn,He:2024pxh,Schacht:2024wgc,Wang:2024ztg}, as it is necessary to explain the baryon-antibaryon asymmetry in the universe.
However, CP violation has only been observed in meson decays, and no definitive evidence of CP violation in baryon decays has been reported in particle physics experiments.
Current observations of CP violation are insufficient to explain the matter-antimatter asymmetry of the universe, underscoring the importance of searching for new sources of CP violation.
In the general theoretical framework of CP violation presented in Ref.~\cite{PDG}, the direct CP violation in two-body decay processes is proportional to the product of the weak phase difference and the strong phase difference: $A_{CP} \propto \sin(\phi_1 - \phi_2) \sin(\delta_1 - \delta_2)$.
Here, $\delta_i$ represents the strong phase, while $\phi_i$ denotes the weak phase originating from the CKM matrix elements.
Under SU(3) symmetry, the strong phase within an SU(3) amplitude corresponding to different channels is identical. With sufficient experimental data, the strong phases of anti-triplet charmed baryon decays can be determined.
Consequently, CP violation can also be incorporated into these processes.

 This paper is organized as follows: In Section II, we present a global analysis of anti-triplet charmed baryon two-body decays using the IRA method.
Section III utilizes the topological diagrams  to gain a comprehensive understanding of SU(3) symmetry.
By determining the previously undetermined parameters from prior work~\cite{Xing:2023dni}, we further numerically investigate the equivalence between the IRA and TDA methods.
Subsequently, the strong phase is incorporated
into the form factor that constitutes the SU(3) amplitude,  with determined values that carry large uncertainties.
The determined strong phases motivate us to explore CPV in these processes.
In Section IV, CPV is expressed using the SU(3) amplitude with contributions from two sources of weak and strong phases under specific scenarios.
The conclusions are presented in Section V.

\section{The global analysis of anti-triplet charmed baryon two-body decays with IRA}\label{se_su3}
Using SU(3) flavor symmetry, the anti-triplet charmed baryon, light baryon octet, and light meson octet can be expressed by $3 \times 3$ matrices $T_{c\bar{3}}$, $T_8$, and $P$, respectively~\cite{Xing:2023dni} as
\begin{eqnarray}
T_{c\bar3}&=&
\begin{pmatrix}
0& \Lambda_c^+ &\Xi_c^+ \\
-\Lambda_c^+ & 0&\Xi_c^0\\
-\Xi_c^+& -\Xi_c^0&0
\end{pmatrix},
P=
\begin{pmatrix}
\frac{\pi^0+\eta_q}{\sqrt{2}}& \pi^+ &K^+ \\
\pi^- & \frac{-\pi^0+\eta_q}{\sqrt{2}}&K^0\\
K^-& \bar{K}^0&\eta_s
\end{pmatrix},\notag\\
T_8&=&
\begin{pmatrix}
\frac{\Sigma^0}{\sqrt{2}}+\frac{\Lambda^0}{\sqrt{6}}& \Sigma^+ &p \\
\Sigma^- & -\frac{\Sigma^0}{\sqrt{2}}+\frac{\Lambda^0}{\sqrt{6}}&n\\
\Xi^-& \Xi^0&-\frac{2\Lambda^0}{\sqrt{6}}
\end{pmatrix}.
\end{eqnarray}
Here the anti-triplet charmed baryon can also be expressed as $(T_{c\bar3})_i=\epsilon^{ijk}(T_{c\bar 3})^{[jk]}=(\Xi_c^0,-\Xi_c^+,\Lambda^+_c)$, and the $\eta_s$ and $\eta_q$ are the mixture of $\eta_1$ and $\eta_8$: $\eta_8=\eta_q/\sqrt{3}-\eta_s\sqrt{2}/\sqrt{3},\quad \eta_1=\eta_q \sqrt{2}/\sqrt{3}+\eta_s/\sqrt{3}$.
To analyze the experimental data, we  consider the physical mixing effects with $\eta^{(\prime)}$  states
\begin{eqnarray}
&&\begin{pmatrix}
\eta \\
\eta^\prime\\
\end{pmatrix}=\begin{pmatrix}
\cos\phi&-\sin\phi \\
\sin\phi&\cos\phi\\
\end{pmatrix}\begin{pmatrix}
\eta_q \\
\eta_s\\
\end{pmatrix},
\end{eqnarray}
where  the physical mixing angles  $\phi=(39.3\pm1.0)^{\circ}$~\cite{Gan:2020aco}.

With these defined matrices, the SU(3)-invariant decay amplitudes for anti-triplet charmed baryon two-body decays can be expressed as
 $a_{15} \times\epsilon_{imn}(T_{c\bar{3}})^{[mn]}(H_{15})^{\{ik\}}_j(\overline{T_8})^j_kP^l_l$ and $a_{6} \times(T_{c\bar{3}})^{[ik]}(H_{\overline{6}})_{\{ij\}}
(\overline{T_8})^j_kP^l_l$.
Here the $H_{15,\bar 6}$ is the decomposed matrix element of the effective weak interaction Hamiltonian. 
In the IRA method, the four-Fermi effective Hamiltonian, such as $c\to s \bar{d} u$, $c\to s \bar{s} u$, or $c\to u \bar{q} q$, is decomposed as $3\otimes \bar 3\otimes 3=3\oplus 3\oplus\bar 6\oplus 15$.
The detailed decomposition formula can be found in Ref.~\cite{Xing:2024nvg}.
The nonzero matrices are $H_{\bar 6}$ and $H_{15}$. 
By enumerating all possible matrix combinations, nine independent amplitudes can be constructed as
follows:
\begin{eqnarray}
\mathcal{M}^{IRA}
&=&a_{15} \times(T_{c\bar{3}})_i(H_{15})^{\{ik\}}_j(\overline{T_8})^j_kP^l_l\notag\\
&+&b_{15} \times(T_{c\bar{3}})_i(H_{15})^{\{ik\}}_j(\overline{T_8})^l_kP^j_l\notag\\
&+&c_{15} \times(T_{c\bar{3}})_i(H_{15})^{\{ik\}}_j(\overline{T_8})^j_lP^l_k\notag\\
&+&d_{15} \times(T_{c\bar{3}})_i(H_{15})^{\{jk\}}_l(\overline{T_8})^l_jP^i_k\notag\\
&+&e_{15} \times(T_{c\bar{3}})_i(H_{15})^{\{jk\}}_l(\overline{T_8})^i_j P^l_k\notag\\
&+&a_{6} \times(T_{c\bar{3}})^{[ik]}(H_{\overline{6}})_{\{ij\}}
(\overline{T_8})^j_kP^l_l\notag\\
&+&b_{6} \times(T_{c\bar{3}})^{[ik]}(H_{\overline{6}})_{\{ij\}}(\overline{T_8})^l_kP^j_l\notag\\
&+&c_{6} \times(T_{c\bar{3}})^{[ik]}(H_{\overline{6}})_{\{ij\}}(\overline{T_8})^j_lP^l_k\notag\\&+&d_{6} \times(T_{c\bar{3}})^{[lk]}(H_{\overline{6}})_{\{ij\}}(\overline{T_8})^i_kP^j_l.\label{su3}
\end{eqnarray}
Here $(a-e)_{15,6}$ means the corresponding SU(3) amplitudes. Using these nine SU(3) amplitudes, all amplitudes of anti-triplet charmed baryon two-body decays can be expressed. The specific formulas for these amplitudes can be found in Ref.~\cite{Xing:2023dni}.

To correctly describe the phenomenological results of anti-triplet charmed baryon two-body decays, each SU(3) amplitude can be expressed by parity-violating form factors $f^{b,c,d}_{6,15}, f^{a(\prime)}$ and parity-conserving form factors $g^{b,c,d}_{6,15},g^{a(\prime)}$ as 
\begin{eqnarray}
  q_6&=&G_F\bar{u}(f^q_6 - g^q_6\gamma_5)u,\quad q=a,b,c,d,\notag\\
  q_{15}&=&G_F\bar{u}(f^q_{15} - g^q_{15}\gamma_5)u,\quad q=a,b,c,d,e.\label{ffira}
\end{eqnarray}

Subsequently, the branching ratios and polarization parameters $\alpha$, $\beta$ and $\gamma$ are expressed as
\begin{eqnarray}
&&\frac{d\Gamma}{d\cos\theta_M}=\frac{G_{F}^{2}|\vec{p}_{B_{n}}|(E_{B_n}+M_{B_{n}})}{8\pi M_{B_c}}(|F|^2+\kappa^2 |G|^2)\notag\\
&&\qquad \qquad \times(1+\alpha \hat\omega_i\cdot\hat p_{B_n} ),\notag\\
&&\alpha=\frac{2\rm{Re}(F*G)\kappa}{(|F|^2+\kappa^2 |G|^2)},\;\;
\beta=\frac{2\mathrm{Im}(F^*G)\kappa}{(|F|^2+\kappa^2 |G|^2)},\notag\\
&&\gamma=\frac{|F|^2-|\kappa G|^2}{(|F|^2+\kappa^2 |G|^2)},\;\;
\kappa=\frac{|\Vec{p}_{B_n}|}{(E_{B_n}+M_{B_n})}.\label{observable}
\end{eqnarray}

Note that most of the experimental data involving $\eta$ and $\eta^\prime$ contain the combination $a_6 - a_{15}$, except for the $\Xi_c^0 \to \Xi^0 \eta$ and $\Xi_c^0 \to \Xi^0 \eta^\prime$ processes, which involve the combination $a_6 + a_{15}$, as shown in Tables II and III of Ref.~\cite{Xing:2023dni}. 
It is expected that the amplitudes $a_6$ and $a_{15}$ will have large uncertainties because only the combination $a_6 - a_{15}$ can be determined precisely. To account for this uncertainty, we redefine the new SU(3) irreducible amplitudes and corresponding form factors as follows:
\begin{eqnarray}
&&f^a=f^a_6-f^a_{15}, \quad  g^a=g^a_6-g^a_{15}\nonumber\\
&& f^{a\prime}=f^a_6+f^a_{15},  \quad g^{a\prime}=g^a_6+g^a_{15}\;.
\end{eqnarray}
The previous experimental data can determine the parameter $a$ with precision, but not $a'$~\cite{Xing:2023dni}. 
Fortunately, the amplitudes from the newly measured processes, $\Xi_c^0\to\Xi^0\eta^{(\prime)}$, depend on $a'$. This means that the corresponding form factors $f^{a\prime}$ and $g^{a\prime}$ can be obtained, which were absent in previous work. Therefore, it is imperative to revisit the global analysis of anti-triplet charmed baryon two-body decays based on these latest results.

\begin{table*}[htbp!]
\caption{Experimental data and fitting results of anti-triplet charmed baryons two-body decays  for two different fits. Case I(II)[III] means the fit results for the real(complex)[new] form factors. }\label{data}
\begin{tabular}{|c|c|c|c|c|c|c|c|c|c|}\hline\hline
\multirow{2}{*}{channel} & \multicolumn{2}{c|}{exp} &\multicolumn{2}{c|}{Case I} &
\multicolumn{2}{c|}{Case II }
&\multicolumn{2}{c|}{Case III }\cr\cline{2-9}
&Br($\%$)& $\alpha$  &Br($\%$)& $\alpha$ &Br($\%$)& $\alpha$ &Br($\%$)& $\alpha$ \\\hline
$\Lambda^{+}_{c}\to p \pi^0 $& $0.0156(75)$
& &$0.0163(60)$ 
& &$0.0158(75)$ & & 0.0141(65)&\\\hline
$\Lambda^{+}_{c}\to p K_S^0 $  &$1.59(7)$  
& $0.2(5)$
&$1.581(47)$ 
&$0.39(14)$ &$1.580(69)$& $-0.05(38)$& 1.584(48) &0.15(24)\\\hline
$\Lambda^{+}_{c}\to p K_L^0 $  &$1.67(7)$  
& 
& $1.689(49)$
& &$1.677(69)$& &1.690(51) &\\\hline
$\Lambda^{+}_{c}\to p\eta $ &$0.158(11)$ 
& &$0.1583(97)$ & &
$0.158(11)$&& 0.154(11) &\\\hline
$\Lambda^{+}_{c}\to p\eta^\prime$& $0.0484(91)$
&  &$0.0484(61)$ &
&$0.0488(91)$& &0.0505(74) &
\\\hline
$\Lambda^{+}_{c}\to \Lambda \pi^+ $  & $1.29(5)$
&$-0.755(6)$
&$1.309(47)$ &$-0.7536(60)$&$1.272(48)$&$-0.7551(60)$ & 1.270(48) &-0.7551(60)\\\hline
$\Lambda^{+}_{c}\to \Sigma^0\pi^+ $ & $1.27(6)$
&$-0.466(18)$&$1.248(46)$ &$-0.472(15)$&$1.245(48)$& $-0.472(15)$ & 1.246(48)& -0.472(15)\\\hline
$\Lambda^{+}_{c}\to \Sigma^{+}\pi^0 $ & $1.24(9)$
&$-0.484(27)$&$1.262(46)$ &$-0.470(15)$&$1.253(48)$&$-0.471(15)$ &1.257(48) &-0.471(15)\\\hline
$\Lambda^{+}_{c}\to \Xi^{0}K^{+} $ &$0.55(7)$
&$0.01(16)$&$0.423(29)$& &$0.555(70)$&0.04(15) & 0.558(68)&0.05(13)\\\hline
$\Lambda^{+}_{c}\to \Lambda^{0}K^{+} $& $0.0642(31)$
&$-0.58(5)$& $0.0639(29)$ & $-0.547(44)$
& $0.0645(31)$ & $-0.585(49)$ &0.0644(31) &-0.588(48)
\\\hline
$\Lambda^{+}_{c}\to \Sigma^{+}\eta $ & $0.32(5)$
&$-0.99(6)$&$0.299(47)$ &$-0.989(29)$&
$0.32(5)$&$-0.985(60)$ &0.338(49) &-0.982(59)\\\hline
$\Lambda^{+}_{c}\to \Sigma^{+}\eta^\prime $  & $0.41(8)$
&$-0.460(67)$&$0.428(66)$& $-0.467(64)$
&$0.407(80)$&$-0.460(67)$ & 0.380(68)&-0.443(64)\\\hline
$\Lambda^{+}_{c}\to \Sigma^{0}K^+ $ & $0.0370(31)$
&$-0.54(20)$&$0.0377(18)$ & $-0.9960(42)$
&$0.0389(25)$& $-0.58(16)$ &0.0386(25) &-0.58(14)\\\hline
$\Lambda^+_c\to n\pi^+$ &  $0.066(13)$
&&$0.0642(23)$&&$0.0766(88)$ & & 0.0765(78)&\\\hline
$\Lambda^+_c\to \Sigma^+K_S^0$ & $0.047(14)$
&&$0.0277(27)$ &&$0.042(13)$& &0.0323(43) &\\\hline
$\Xi^{+}_{c}\to \Xi^{0}\pi^+ $ & $1.6(8)$
&&$0.875(77)$ &
&$1.94(59)$& & 2.02(32)&\\\hline
$\Xi^{+}_{c}\to pK_S $
& $0.0716(325)$
&&0.144(14) &
&& & &\\\hline
$\Xi^{+}_{c}\to \Lambda^{0}\pi^+ $ & $0.0452(209)$
&&0.0218(32) &
&& & &\\\hline
$\Xi^{+}_{c}\to \Sigma^{0}\pi^+ $ & $0.120(55)$
&&0.3161(88) &
&& & &\\\hline
$\frac{\Xi^{+}_{c}\to p K_S}{\Xi^{+}_{c}\to \Lambda^{0}\pi^+} $ & 1.58(21)
&& &
&1.62(20)& &1.51(18) &\\\hline
$\frac{\Xi^{+}_{c}\to \Lambda^{0}\pi^+}{\Xi^{+}_{c}\to \Sigma^{0}\pi^+} $ & 0.378(52)
&& &
&0.375(45)& &0.352(39) &\\\hline
$\Xi^{0}_{c}\to \Lambda K_S^0 $ & $0.32(6)$
&&$0.225(34)$&
&$0.324(60)$& &0.345(57) &\\\hline
$\Xi^{0}_{c}\to \Xi^- \pi^+ $ & $1.43(27)$
&$-0.640(51)$&$1.16(18)$& $-0.709(45)$
&$1.18(19)$& $-0.640(51)$ &1.17(18) &-0.640(51)\\\hline
$\Xi^{0}_{c}\to \Xi^- K^+ $ & $0.039(11)$
&&$0.0515(80)$& &$0.0491(82)$& & 0.0508(81)&\\\hline
$\Xi_c^0\to\Sigma^0 K^0_S$ &$0.054(16)$
&&$0.054(16)$ &&$0.055(16)$& &0.056(16) &\\\hline
$\Xi_c^0\to\Sigma^+ K^-$ &$0.18(4)$
&&$0.195(39)$ && $0.181(40)$ & & 0.183(40)&
\\\hline
$\Xi_c^0\to\Xi^0 \pi^0$ & $0.69(14)$ 
& $-0.90(28)$ &$0.152(48)$ 
&$-0.45(13)$&$0.68(14)$&$-0.87(28)$ &0.60(11) & -0.87(26)\\\hline
$\Xi_c^0\to\Xi^0 \eta$ & $0.16(4)$
&&$0.16(4)$ &&$0.16(4)$& &0.159(40) &\\\hline
$\Xi_c^0\to\Xi^0 \eta'$ & $0.12(4)$
&  &$0.12(4)$& &$0.12(4)$& & 0.126(40)&\\\hline
\hline  
\end{tabular}
\end{table*}

By assuming real form factors for simplicity, we perform fits using the updated data from Eq.~\ref{belle} and the latest PDG~\cite{PDG} in Table~\ref{data} using the nonlinear least-$\chi^2$ method~\cite{Fit}.
Here, we disregard the BESIII result $\alpha(\Lambda_c^+ \to \Xi^0 K^+) = 0.01 \pm 0.16 \pm 0.03$~\cite{BESIII:2023wrw}, which has a large error compared to the central value and is inconsistent with our previous work~\cite{Xing:2023dni}.
In our preliminary analysis, the channels $\Xi_c^0 \to \Xi^0 \pi^0$ and $\Xi^+_c\to \Sigma^0\pi^+$ contribute the most to the $\chi^2$. 
The situation we encountered of $\Xi_c^0 \to \Xi^0 \pi^0$ is same with previous work~\cite{He:2024pxh}, while the measurement of the branching ratio for $\Xi^+_c\to \Sigma^0\pi^0$ is in conflict with previous predictions~\cite{Xing:2023dni,He:2024pxh,Cheng:2024lsn}.

Therefore, we attempt to exclude these measurements from the global fit and present the fit results (Case I) in Table~\ref{table2}. 
Note that the new measurements of the three channels $\Xi^+_c\to p K_S$, $\Xi^+_c\to \Lambda^0\pi^+$, and $\Xi^+_c\to \Sigma^0\pi^+$ are obtained through the ratio $Br(\Xi^+_c\to T_8 P)/Br(\Xi_c^+\to \Xi^-2\pi^+)$.
The large uncertainty of $Br(\Xi_c^+\to \Xi^-2\pi^+)=(2.9\pm1.3)\%$ will increase the uncertainty in these three measurements naturally. 
However, the ratios $Br(\Xi^+_c\to p K_S)/Br(\Xi^+_c\to \Lambda^0\pi^+)$ and $Br(\Xi^+_c\to \Lambda^0\pi^+)/Br(\Xi^+_c\to \Sigma^0\pi^+)$ can effectively eliminate the uncertainty from $Br(\Xi_c^+\to \Xi^-2\pi^+)$. In the fit of Case I, we find that these predicted ratios differ significantly from the experimental data in Table.~\ref{data}.
Based on the fitted results, $Br(\Xi^{+}_{c}\to pK_S)=0.144(14)$,  $Br(\Xi^+_c\to \Lambda^0\pi^+)=0.0218(32)$,  $Br(\Xi^+_c\to \Sigma^0\pi^+)=0.3161(88)$,
 these ratios are predicted as
\begin{eqnarray}
&&\frac{Br(\Xi^+_c\to p K_s)}{Br(\Xi^+_c\to \Lambda^0\pi^+)}=6.6\pm 1.2, \nonumber\\ 
&&\frac{Br(\Xi^+_c\to \Lambda^0\pi^+)}{Br(\Xi^+_c\to \Sigma^0\pi^+)}=0.069\pm 0.010.
\end{eqnarray}
This also suggests that the assumption of a real form factor no longer aligns with the current experimental data.

\begin{table*}[htbp!]
\caption{The fit results for the real (Case I), complex (Case II), new (Case III) and TDA form factors. }\label{table2}
\begin{tabular}
{|c|c|c|c|c|c|c|c|c|c|}\hline\hline
form factors 
&\multicolumn{5}{c|}{ Case I ($\chi^{2}$/d.o.f=28.49/18=1.58)} 
\\\hline
\multirow{2}{*}{vector(f) } 
 & $f^a=0.0103(27)$ & $f^{b}_{6}=0.0193(47)$
 & $f^{c}_{6}= 0.0234(42)$ 
&$f^d_6=-0.0090(40)$ 
& $f^{a\prime}=0.0007(73)$ \\\cline{2-6}
&$f^{b}_{15}=-0.0103(25)$
& $f^{c}_{15}=0.0065(43)$
 & 
 $f^d_{15}=-0.0154(22)$ &
 $f^e_{15}= 0.0530(41)$ &\\\cline{2-6} 
\hline
\multirow{2}{*}{axial-vector(g) } &
$g^a=-0.0295(79)$ & $g^{b}_{6}=-0.1767(57)$
& $g^{c}_{6}= 0.0896(90)$
&$g^d_6=-0.0614(73)$
& $g^{a\prime}=0.025(44)$ \\\cline{2-6} 
& $g^{b}_{15}=0.0739(49)$
& $g^{c}_{15}=0.0024(89)$
& 
$g^d_{15}=-0.0150(59)$ &
$g^e_{15}= 0.0170(36)$&\\\cline{2-6}
\hline\hline
\multirow{2}{*}{form factors } &\multicolumn{5}{c|}{ Case II ($\chi^{2}$/d.o.f=4.07/3=1.36)}  \cr\cline{2-6}&\multicolumn{3}{c|}{absolute value} & \multicolumn{2}{c|}{strong phase}\\\cline{1-6}
\multirow{4}{*}{vector(f) } 
&$f^a=0.039(29)$  
&$f^b_6=0.007(48)$ 
 &$f^{c}_{6}= 0.022(9.63)$
 & $\delta f^b_{6}=-2.356(1.179)$ 
 &  $\delta f^c_{b}=2.035(209)$
\\\cline{2-6}
 &$f^d_6=0.007(38)$ &$f^{a\prime}=0.014(13.97)$
& 
& $\delta f^d_{6}=2.467(1.435)$ 
 &  $\delta f^{a\prime}=-2.304(3.143)$ 
 \\\cline{2-6}
&$f^{b}_{15}=0.032(79)$
& $f^{c}_{15}=0.004(9.63)$  
& 
&  $\delta f^b_{15}=1.190(1.916)$
&  $\delta f^c_{15}=-1.816(1.621)$\\\cline{2-6}
&$f^d_{15}=0.023(33)$
& $f^e_{15}= 0.045(158)$ 
&
&  $\delta f^d_{15}=-1.347(935)$
& $\delta f^e_{15}=-3.141(6.236)$
\\\cline{2-6}
\hline
\multirow{5}{*}{axial-vector(g) } 
&$g^a=0.120(368)$ 
&$g^{b}_{6}=0.110(284)$ 
& 
 & $\delta g^a=-3.125(1.665)$ 
 & $\delta g_6^b=0.027(642)$ \\\cline{2-6}
& $g^{c}_{6}= 0.021(14.47)$
&$g^d_6=0.042(121)$ 
&  
 & $\delta g_6^c=1.907(3.802)$
& $\delta g_{6}^d=-0.802(2.562)$
\\\cline{2-6}
& $g^{a\prime}=0.099(707)$
& 
&
 & $\delta g^{a\prime}=-1.153(342)$
& 
\\\cline{2-6}
&$g^{b}_{15}=0.17(425)$
& $g^{c}_{15}=0.045(14.47)$ 
& 
& $\delta g_{15}^b=2.061(2.209)$
&$\delta g_{15}^c=1.107(1.444)$\\\cline{2-6}
&$g^d_{15}=0.056(175)$
& $g^e_{15}= 0.008(126)$ & 
& $\delta g_{15}^d=-3.053(508)$ 
&$\delta g_{15}^e=1.546(4.42)$\\\cline{2-6}
\hline\hline
\multirow{2}{*}{form factors } &\multicolumn{5}{c|}{ Case III ($\chi^{2}$/d.o.f=7.03/7=1.004)}  \cr\cline{2-6}&\multicolumn{3}{c|}{real part} & \multicolumn{2}{c|}{imaginary part}\\\cline{1-6}
\multirow{4}{*}{vector(f) } 
&$f^a=-0.0289(61)$  
&$f^c_6=0.010(10)$ 
 &$f^{d}_{6}=-0.008(17)$
 & $f_{T_1}=-0.0176(37)$ 
 &  $f_{T_4}=-0.0019(81)$
\\\cline{2-6}
 &$f^{a\prime}=-0.01(50)$ &${\rm Re} (\textbf{f}_b^S)=0.0003(114)$
& ${\rm Re}(\textbf{f}_b^P)=0.000051(40)$
& ${\rm Im}(\textbf{f}_b^S)=-0.0105(36)$ 
 &  ${\rm Im}(\textbf{f}_b^P)=-0.000049(80)$ 
 \\\cline{2-6}
&$f^{b}_{15}=0.0326(72)$
& $f^{c}_{15}=-0.0038(59)$  
& $f^{d}_{15}=-0.004(16)$ 
&  $f_{T_6}=-0.0192(71)$
&  $f_{T_7}=0.031(12)$\\\cline{2-6}
& $f^e_{15}= -0.0338(76)$ 
& 
&
&  
& 
\\\cline{2-6}
\hline
\multirow{4}{*}{axial-vector(g) } 
&$g^a=-0.014(18)$ 
&$g^{b}_{6}=0.083(41)$ 
& $g^{c}_{6}=-0.023(39)$
 & $ g_{T_1}=-0.016(18)$ 
 & $g_{T_3}=-0.08(1.52)$ \\\cline{2-6}
& $g^{d}_{6}= 0.031(21)$
&$g^{a\prime}=0.05(49)$ 
&  
 & $g_{T_4}=0.068(36)$
& $g_{T_6}=0.031(26)$
\\\cline{2-6}
&$g^{b}_{15}=-0.153(31)$
& $g^{c}_{15}=-0.006(26)$ 
& $g^{d}_{15}=-0.029(13)$
& $ g_{T_7}=-0.039(47)$
&\\\cline{2-6}
&$g^e_{15}= -0.016(20)$
&  & 
&  
&\\\cline{2-6}
\hline\hline
 form factors  &\multicolumn{5}{c|}{ TDA 
 } 
\\\hline
\multirow{8}{*}{vector(f) }
& $T_1$ & $T_2$ & $T_3$ &$T_4$ &$T_5$ \\\cline{2-6}
&$f_{1}=0.0813(47)$ &$f_{2}=0.0247(58)$ 
& $f_3=-0.0052(14)$ 
& $f_{4}=-0.0084(11)$ 
& $f_{7}= 0.0149(41)$
\\ 
& $f_{15}=0.0406(24)$
& $f_{16}=0.0123(29)$
& $f_{5}= 0.0003(36)$ 
&$f_6=-0.0527(42)$
&$f_8=-0.0156(40)$
\\
&  
& 
&  $f_9=0.0055(38)$
&$f_{10}=-0.0443(44)$
&$f_{11}=-0.0305(59)$ \\ \cline{2-6}
 &  $T_6$ &  $T_7$& & &\\\cline{2-6} 
&  $f_{12}=0.0234(42)$
&  $f_{17}=0.0271(18)$
& & &\\ 
 & $f_{13}=-0.0252(33)$ 
 & $f_{18}= 0.0361(41)$
  & & &\\
  & $f_{14}=-0.0486(36)$
  &$f_{19}=-0.0090(40)$& & &\\\cline{2-5} 
\hline
\multirow{8}{*}{axial-vector(g) } 
& $T_1$ & $T_2$ & $T_3$ &$T_4$ &$T_5$ \\\cline{2-6}
&$g_1=-0.0982(86)$ 
& $g_{2}=0.1323(77)$
&$g_3=0.0147(39)$ & $g_{4}=-0.0436(31)$
& $g_{7}= 0.0460(84)$
\\
& $g_{15}=-0.0491(43)$
& $g_{16}=0.0662(38)$
& $g_{5}= 0.012(22)$
&$g_6=0.026(12)$ 
& $g_8=-0.058(14)$\\
& 
& 
& $g_9=-0.002(22)$ 
&$g_{10}=0.070(11)$ 
&$g_{11}=-0.104(15)$\\\cline{2-6} 
 & $T_6$ & $T_7$  & && \\\cline{2-6}
&  $g_{12}=0.0896(90)$
& $g_{17}=-0.0591(37)$
& &&\\
& $g_{13}=0.052(14)$
& $g_{18}= 0.0023(58)$
 & & &
\\
& $g_{14}=-0.037(13)$
&$g_{19}=-0.0614(73)$
& & &\\\cline{2-5}
\hline\hline
\end{tabular}
\end{table*}

In our result (Case I), we predict $\alpha(\Lambda_c^+ \to \Sigma^0 K^+) = -0.9960 \pm 0.0042$, which deviates by $2\sigma$ from the experimental data $\alpha(\Lambda_c \to \Sigma^0 K^+)_{\text{exp}} = -0.54 \pm 0.20$. The value $\alpha(\Lambda_c^+ \to \Xi^0 K^+)=0.957\pm0.018$ also conflicts with the experimental measurement $\alpha(\Lambda_c^+ \to \Xi^0 K^+)_{\text{exp}} = 0.01 \pm 0.16 \pm 0.03$.
These conflicting polarization parameter data indicate that the assumption of real form factors is insufficient, as $\alpha$ is proportional to $\rm{Re}(F*G)$ which is highly dependent on the complex phase of the form factors $F$ and $G$.


The above analysis demonstrates that the strong phase of the form factors defined in Eq.~\ref{ffira} is essential. We can decompose these form factors into their absolute values and phases as follows:
\begin{eqnarray}
f^q_{6}&=&|f^q_6|e^{i\delta f^q_6},\;\;g^q_{6}=|g^q_6|e^{i\delta g^q_6},q=a,b,c,d,\notag\\
f^q_{15}&=&|f^q_{15}|e^{i\delta f^q_{15}},g^q_{15}=|g^q_{15}|e^{i\delta g^q_{15}},q=a,b,c,d,e.
\end{eqnarray}

Before conducting our global analysis, the degrees of freedom can be calculated using the formula $d.o.f = N - M + 1$, where $N = 37$ represents the number of experimental data points, and $M$ is the number of parameters used in the analysis.
Note that  the ratios $Br(\Xi^+_c\to p K_S)/Br(\Xi^+_c\to \Lambda^0\pi^+)$ and $Br(\Xi^+_c\to \Lambda^0\pi^+)/Br(\Xi^+_c\to \Sigma^0\pi^+)$ are adopted to reduce the uncertainty rather than the respective branching ratios.
Since we introduce 18 form factors in our analysis, and each form factor consists of its absolute value and phase, the total number of parameters, $M$, can be calculated as $M = 18 \times 2 - 1$, after subtracting the global phase and setting $\delta f^a_6 = 0$. 
The degrees of freedom (d.o.f) in this fit is $37 - 35 + 1 = 3$. The small d.o.f suggests that the experimental data is insufficient to precisely determine the strong phase, leading to large uncertainties in the fit results.

After including all 35 parameters in our global fit, we derive the numerical results of form factors presented in Table~\ref{table2} (case II), with a $\chi^2/d.o.f = 1.36$. 
However, we find that the complex form still struggles to simultaneously explain the new measurements $\Xi^+_c\to p K_S$, $\Xi^+_c\to \Lambda^0\pi^+$, and $\Xi^+_c\to \Sigma^0\pi^+$. While their individual ratios can be fitted accurately,  more precise experimental data on  $Br(\Xi_c^+\to \Xi^-2\pi^+)$ is required to obtain the accurate respective branching ratios in the future.

The low $\chi^2/d.o.f$ indicates that the SU(3) symmetry is a reliable and effective symmetry, as it fits the experimental data well.
The predictions (case II) calculated using the fitted complex 18 form factors are shown in Table.~\ref{tableeta} and Table.~\ref{tableccomplex}.
One can find that the added phase of the form factor can perfectly explain the anomaly in the polarization parameter $\alpha(\Lambda_c \to \Sigma^0 K^+)$ and branching ratios $Br(\Xi_c^0 \to \Xi^0 \pi^0)$.
However, as expected, the uncertainty in the absolute value and phase is very large, which reduces the predictive power of our results. Therefore, we still look forward to more experimental data to reduce the uncertainty in our results.

\section{The comprehensive analysis of  anti-triplet charmed baryon two body decays with the equivalence of TDA and IRA }\label{iratda}

In the above analysis, all nine IRA amplitudes and the 18 associated form factors are determined based on the key experimental results measured by Belle and Belle II, as shown in Table~\ref{table2}. 
This facilitates a detailed and comprehensive study of anti-triplet charmed baryon two-body decays.
A comprehensive and intuitive analysis can be performed by introducing the topological diagrammatic approach (TDA), which provides clearer physical insights.

\begin{figure}
    \centering
    \includegraphics[width=0.94\columnwidth]{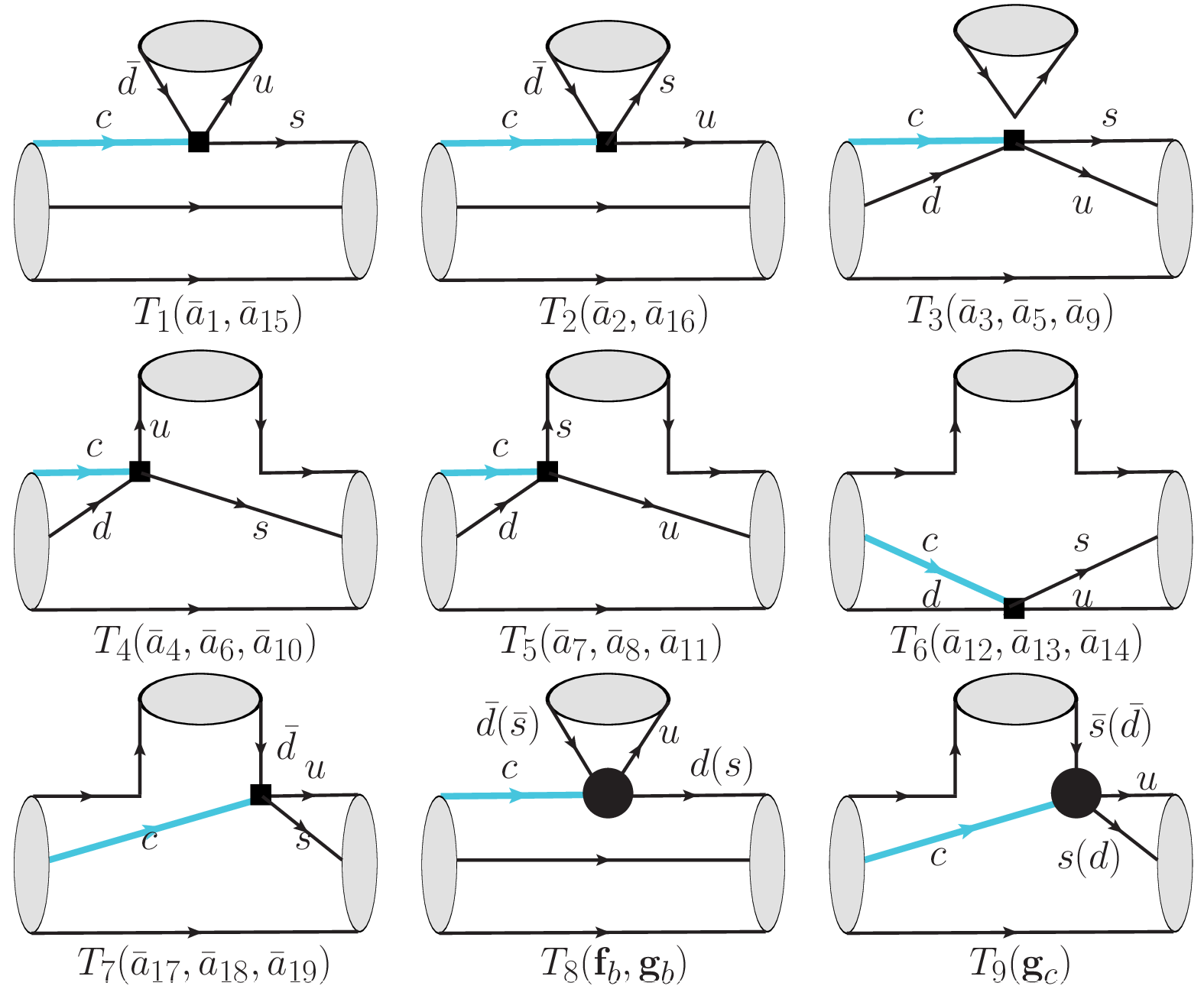}
    \caption{Topology diagrams for the charmed baryon two body decays. Here $T_{1-7}$   corresponding to the TDA amplitude in Eq.\ref{TDA}.
 The Cabibbo-suppressed $c\to u\bar d d (\bar s s)$ and doubly-suppressed $c\to u\bar s d$ can be obtained similarly. 
 $T_{8,9}$ describe the new effects in the Cabibbo-suppressed $c\to u\bar d d (\bar s s)$  induced by $\textbf{f}_b$, $\textbf{g}_b$ and  $\textbf{g}_c$ respectively.}
    \label{fig1}
\end{figure}

In fact, the two different methods (IRA and TDA) are mutually equivalent, as thoroughly analyzed in Refs.~\cite{He:2018joe,Hsiao:2021nsc,Zhong:2024qqs}. In our work, we adopt the same notation for the TDA amplitudes, denoted as ${\bar{a}_{1\sim19}}$, following Ref.~\cite{He:2018joe} as
\begin{eqnarray}
&&\mathcal{M}^{TDA}=
\bar{a}_1T_{c\bar{3}}^{[ij]}H_m^{kl}(\overline{T}_8)_{ijk}P_l^m\notag\\
&&+\bar{a}_2T_{c\bar{3}}^{[ij]}H_m^{kl}(\overline{T}_8)_{ijl}P_k^m
+\bar{a}_3T_{c\bar{3}}^{[ij]}H_i^{kl}(\overline{T}_8)_{jkl}P_m^m\notag\\
&&+\bar{a}_4T_{c\bar{3}}^{[ij]}H_i^{kl}(\overline{T}_8)_{jkm}P_l^m  
+\bar{a}_5T_{c\bar{3}}^{[ij]}H_i^{kl}(\overline{T}_8)_{jlk}P_m^m\notag\\
&&+\bar{a}_6T_{c\bar{3}}^{[ij]}H_i^{kl}(\overline{T}_8)_{jmk}P_l^m
+\bar{a}_7T_{c\bar{3}}^{[ij]}H_i^{kl}(\overline{T}_8)_{jlm}P_k^m\notag\\
&&+\bar{a}_8T_{c\bar{3}}^{[ij]}H_i^{kl}(\overline{T}_8)_{jml}P_k^m 
+\bar{a}_9T_{c\bar{3}}^{[ij]}H_i^{kl}(\overline{T}_8)_{klj}P_m^m\notag
\end{eqnarray}
\begin{eqnarray}
&&+\bar{a}_{10}T_{c\bar{3}}^{[ij]}H_i^{kl}(\overline{T}_8)_{kmj}P_l^m
+\bar{a}_{11}T_{c\bar{3}}^{[ij]}H_i^{kl}(\overline{T}_8)_{lmj}P_k^m \notag\\
&&+\bar{a}_{12}T_{c\bar{3}}^{[ij]}H_i^{kl}(\overline{T}_8)_{klm}P_j^m
+\bar{a}_{13}T_{c\bar{3}}^{[ij]}H_i^{kl}(\overline{T}_8)_{kml}P_j^m\notag\\
&&+\bar{a}_{14}T_{c\bar{3}}^{[ij]}H_i^{kl}(\overline{T}_8)_{lmk}P_j^m
+\bar{a}_{15}T_{c\bar{3}}^{[ij]}H_m^{kl}(\overline{T}_8)_{ikj}P_l^m\notag\\
&&+\bar{a}_{16}T_{c\bar{3}}^{[ij]}H_m^{kl}(\overline{T}_8)_{ilj}P^m_k
+\bar{a}_{17}T_{c\bar{3}}^{[ij]}H_m^{kl}(\overline{T}_8)_{ikl}P_j^m\notag\\
&&+\bar{a}_{18}T_{c\bar{3}}^{[ij]}H_m^{kl}(\overline{T}_8)_{ilk}P_j^m
+\bar{a}_{19}T_{c\bar{3}}^{[ij]}H_m^{kl}(\overline{T}_8)_{klj}P_i^m.
\label{TDA}\end{eqnarray}
The corresponding topological diagrams are illustrated in Fig.~\ref{fig1}($T_1\sim T_7$).

Using the definition $(T_8)_{ijk}=\epsilon_{ijl}(\bar{T_8})^l_k$, the $H^{ij}_k$ in TDA can be decomposed into IRA forms as 
\begin{eqnarray}
H^{ij}_k=\frac{1}{2}\bigg[ (H_{15})^{ij}_k+\frac{1}{2}\epsilon^{ijl}(H_{\bar 6})_{kl}\bigg]\;.    
\end{eqnarray}
The transformation from TDA to IRA occurs naturally with the following relations for the parameters:
\begin{eqnarray}
&&\overline{a}_1=b_{6}-d_{6}+e_{15},\quad\overline{a}_{2}=d_{6}-b_{6}+e_{15},\quad\overline{a}_{3}=-\frac{a}{2},\notag\\
&&\overline{a}_{4}=\frac{1}{2}(-c_{6}+c_{15}),
\quad\overline{a}_{5}=\frac{1}{2}a^{\prime},\quad\overline{a}_{7}=\frac{1}{2}(c_{6}+c_{15}),\notag\\
&&\overline{a}_{6}=\frac{1}{2}(-b_{6}-c_{6}-e_{15}+d_{15})+\frac{1}{4}(a+a^{\prime}),\notag\\
&&\overline{a}_{8}=\frac{1}{2}(b_{6}+c_{6}+d_{15}-e_{15})-\frac{1}{4}(a+a^{\prime}),\notag\\
&&\overline{a}_{9}=\frac{1}{2}(a+a^{\prime}),\quad \overline{a}_{12}=c_6,
\quad \overline{a}_{15}=\frac{1}{2}(b_6-d_6+e_{15}),
\notag\\
&&\overline{a}_{10}=\frac{1}{2}(-b_{6}-c_{15}+d_{15}-e_{15})+\frac{1}{4}(a+a^{\prime}),\notag\\
&&\overline{a}_{11}=\frac{1}{2}(b_{6}-c_{15}+d_{15}-e_{15})-\frac{1}{4}(a+a^{\prime})),\notag\\
&&\overline{a}_{13}=\frac{1}{2}(-e_{15}+b_{15}+c_{6}+d_{15})-\frac{1}{4}(a'-a)\notag\\
&&\overline{a}_{14}=\frac{1}{2}(-e_{15}+b_{15}-c_{6}+d_{15})-\frac{1}{4}(a'-a)\notag\\
&&\overline{a}_{16}=\frac{1}{2}(-b_{6}+d_{6}+e_{15}),\quad\overline{a}_{17}=\frac{1}{2}(d_6+e_{15}-b_{15}),\notag\\
&&\overline{a}_{18}=\frac{1}{2}(-d_6+e_{15}-b_{15}),\quad \overline{a}_{19}=d_6.\label{equal}
\end{eqnarray}

By using the definition $\bar a_i=G_F\bar{u}(f_i - g_i\gamma_5)u$ with $i={1\sim 19}$, the numerical results of the form factors for TDA are presented in the lower panel of Table.~\ref{table2}.
Since the form factors fitted in case II have significant uncertainty, we rely on the form factors fitted in case I for the numerical analysis.

Note that although diagrams $T_1$ to $T_2$ and $T_4$ to $T_5$ share similar topological structures, they differ in TDA due to the distinct contributions from operators $O_1$ and $O_2$.
As demonstrated in Ref.~\cite{He:2018joe}, each topological diagram in Fig.~\ref{fig1} corresponds to more than one TDA amplitude. Therefore, it is assumed that the absolute values of the TDA amplitudes with the same topological diagram should be equal.
 By defining the form factor $\mathcal{M}_{T_i}=G_F\bar{u}(f_{T_i} - g_{T_i}\gamma_5)u,\; i=1\sim7$, one obtains the relations as
 \begin{eqnarray}\label{AT}
&&|A_{1,15}|=A_{T_1},\quad|A_{2,16}|=A_{T_2},
\quad |A_{3,5,9}|=A_{T_3},\notag\\
&&|A_{4,6,10}|=A_{T_4},\quad
|A_{7,8,11}|=A_{T_5},\notag\\
&&|A_{12,13,14}|=A_{T_6},
\quad|A_{17,18,19}|=A_{T_7},\; A=f,g.\label{tdaff}
\end{eqnarray}

The assumptions are approximately supported by the numerical results within $1\sigma$ in Table.~\ref{table2}, which shows that the diagrams $T_{1,2}$ actually give the main contribution.
However, some form factors contradict the assumption, 
such as   $|f_{T_{17,18}}|-|f_{19}|\sim 0.02$ and $|g_{17,19}|-|g_{18}|\sim 0.05$. 
To focus on the largest effects, we  ignore the discrepancy involving $f_{19}$ and primarily consider $|g_{17,19}|>|g_{18}|$ in the subsequent analysis and this implies that there are other effects in the topology diagrams.
Besides, since the difference of $T_1-T_2$ and $T_4-T_5$ contribution come  from the operator $O_1$ and $O_2$, one can expect that $A_{T_1}/A_{T_2}\approx A_{T_4}/A_{T_5}$. However, our numerical results conflict with it. We find the difference of contribution of diagram $T_1$ and $T_2$ is as large as $|f_{1,15}|-|f_{2,16}|\sim 0.06$, $|g_{T_2}|-|g_{T_1}|\sim 0.04$. The contributions of diagrams $T_4$ and $T_5$ are consistent within a 2$\sigma$ standard deviation. This conflict make it reasonable to introduce a new form factor to account for the difference between $T_1$ and $T_2$, while assuming $A_{T_4} = A_{T_5}$, where $A=f,g$.

For exploring the possible new effects, one defines the form factors $\textbf{f}_b$ and $\textbf{g}_{b,c}$
 to absorb these effect as as $\textbf{f}_b=f_{T_1}-f_{T_2}$, 
 $\textbf{g}_b=g_{T_1}-g_{T_2}$
 and $\textbf{g}_c=-g_{18}-g_{T_7}$.
 The corresponding topology diagrams $T_{8,9}$ are given in Fig.~\ref{fig1}.
Then the IRA form factors are expressed by TDA form factors and  $\textbf{f}_b,\textbf{g}_{b,c}$ as
\begin{eqnarray}\label{iratotda}
&&f^a=-\frac{f_{T_6}}{2},\quad f^{a\prime}=\frac{1}{2}f_{T_6}-f_{T_3}, 
\quad f^b_6=\frac{3}{2}\textbf{f}_b,\notag\\
&&f^b_{15}=-f_{T_6}-f_{T_7},\quad f^c_6=f_{T_6}+f_{T_4},\quad f^c_{15}=f_{T_4},\notag\\
&&f^d_6=-\frac{3}{2}\textbf{f}_b-f_{T_7},\quad f^d_{15}=-2f_{T_4}-f_{T_6},\notag\\
&&f^e_{15}=-\frac{3}{2}\textbf{f}_b+3f_{T_1}+2f_{T_4}+f_{T_6}+f_{T_7},\notag  
\\
&& g^a=g_{T_4}+\frac{g_{T_6}}{2},\quad g^{a\prime}=g_{T_4}+g_{T_3}-\frac{g_{T_6}}{2},\;g^c_{15}=0\notag\\
&& g^b_6=-3g_{T_1}+\frac{3}{2}\textbf{g}_b-2g_{T_4},
\quad g^b_{15}=\frac{\textbf{g}_{c}}{2}+g_{T_6}+g_{T_7},\notag\\
&&g^c_6=g_{T_6},\quad g^d_6=3g_{T_1}-\frac{3}{2}\textbf{g}_b-g_{T_7}+\frac{\textbf{g}_c}{2},\quad g^d_{15}=g_{T_6},\notag
\\
&&  g^e_{15}=-g_{T_6}-g_{T_7}-\frac{\textbf{g}_{c}}{2}-\frac{3}{2}\textbf{g}_b.
\end{eqnarray}
By adding the new form factor, the TDA form factor and IRA form factor become numerically equivalent.
For example, in Eq.~\ref{iratotda}, 
without the new form factor, the IRA form factors $f^b_6$ and $g^c_{15}$ should be zero. Although the IRA form factor $g^c_{15}=0.0024(89)$ we fitted is close to zero, $f^b_6=0.0193(47)$ conflicts with our equivalence analysis. With the help of the new form factor $\textbf{f}_b$, our fitted results align with the equivalence analysis above.

Although the new form factor we added can explain the equivalence of TDA and IRA amplitudes, the origin of these new form factors still needs further investigation.
Since the diagrams $T_8$ and $T_9$, corresponding to the new form factors, share the same topological structure as the diagrams induced by $O_{1,2}$ in TDA diagrams $T_{1-7}$, it is reasonable to conclude that these diagrams $T_{8,9}$ arise from SU(3) symmetry breaking.
It is also evident that the penguin operators $O_{3-6}$ can contribute to the diagrams $T_1$ and $T_7$, but these contributions cannot be directly accounted for in the nine IRA amplitudes outlined in Eq.~\ref{su3}.
Thus, the penguin operators can also contribute to the new form factors we introduced. In addition to these two possible sources, various other effects may also contribute to the formation of these new form factors. These contributions will be explored in detail in the subsequent section.

\begin{table*}[htbp!]
\caption{The predicted values for branching ratios, polarization parameters ($\alpha,\beta,\gamma$) and CP violation with the final states $\eta^{(\prime)}$ for  different fits. Case I(II)[III] means the fit results for the real(complex)[new] form factors.
}\label{tableeta}
\begin{tabular}{|c|c|c|c|c|c|c|c|c|c|c|c|c|c|c|c|c|c}\hline\hline
\multirow{2}{*}{channel} &  \multicolumn{2}{c|}{ Case I}&  \multicolumn{4}{c|}{ Case II} &\multicolumn{1}{c|}{ Case III}\cr\cline{2-8}   & 
Br($\%$)& $\alpha$ & 
Br($\%$)& $\alpha$ & $\beta$ & $\gamma$  
&CPV 
 \\\hline \hline
$\Lambda^{+}_{c}\to \Sigma^{+}  \eta $ &$0.299(47)$&$-0.989(29)$&$0.32(5)$ &$-0.985(60)$&$-0.1(1.3)$&$0.08(2.33)$ 
&\\\hline
$\Lambda^{+}_{c}\to \Sigma^{+}  \eta^\prime $ & $0.428(66)$&$-0.467(64)$& $0.407(80)$&$-0.460(67)$&$-0.4(1.9)$&$0.8(1.1)$ 
&\\\hline
$\Lambda^{+}_{c}\to p  \eta $  &$0.1583(97)$&$0.865(94)$    &$0.158(11)$ &$0.3(5)$&$-0.5(2.5)$ &$-0.8(1.4)$  
&-0.00058(88)\\\hline
$\Lambda^{+}_{c}\to p  \eta^\prime $         & $0.0484(61)$&$-0.992(15)$&$0.0488(91)$ &$-0.28(60)$&$-0.85(76)$ &$0.4(1.6)$  
&0.0017(30)\\\hline
$\Xi^{+}_{c}\to \Sigma^{+} \eta $   & $0.114(14)$&$0.88(15)$ & $0.164(76)$&$0.7(2.1)$&$-0.7(1.8)$&$0.07(2.43)$  
&-0.0008(17)\\\hline
$\Xi^{+}_{c}\to \Sigma^{+}  \eta^\prime $   & $0.118(17)$&$-0.414(72)$ & $0.060(48)$& $-0.9(1.2)$&$-0.4(4.1)$&$0.2(3.0)$  
&0.0016(26)\\\hline
$\Xi^{+}_{c}\to p  \eta $ &$0.00839(64)$&$-0.058(63)$		&$0.012(11)$ &$-0.68(73)$&$-0.69(87)$&$-0.26(78)$  
&\\\hline
$\Xi^{+}_{c}\to p  \eta^\prime $ 		&$0.0092(10)$&$-0.993(11)$&$0.0060(47)$ &$-0.68(58)$&$-0.73(98)$&$0.1(2.9)$  
&\\\hline
$\Xi^{0}_{c}\to \Xi^{0}  \eta $ &$0.163(29)$&$0.98(16)$ &  $0.1(46.3)$&$-0.08(496.29)$&$0.7(211.9)$& $0.7(242.3)$ 
&\\\hline
$\Xi^{0}_{c}\to \Xi^{0}  \eta^\prime $ &$0.116(33)$& $0.89(46)$&$0.09(29.53)$ &$-0.5(145.5)$&$-0.2(756.2)$&$0.8(133.0)$ 
&\\\hline
$\Xi^{0}_{c}\to \Sigma^{0}  \eta $    &$0.0200(48)$& $0.83(22)$&$0.01(2.40)$&$0.5(274.5)$&$0.7(265.0)$& $0.5(196.4)$  
&-0.0007(43)\\\hline
$\Xi^{0}_{c}\to \Sigma^{0}  \eta^\prime$    &$0.0068(16)$&$0.67(68)$&$0.006(4.432)$ &$-0.8(374.1)$&$0.2(545.0)$&$0.6(373.0)$ 
&0.0015(50)\\\hline
$\Xi^{0}_{c}\to \Lambda  \eta $  &$0.0136(40)$& $0.94(14)$ & $0.03(2.09)$&$-0.4(187.5)$&$0.9(66.8)$& $0.06(62.55)$  
&-0.0018(91)\\\hline
$\Xi^{0}_{c}\to \Lambda  \eta^\prime$  &$0.0051(53)$& $0.94(70)$ & $0.02(14.18)$&$-0.05(346.33)$&$-0.7(213.3)$& $0.7(222.3)$ 
&0.0004(1578)\\\hline
$\Xi^{0}_{c}\to n \eta $  &$0.00053(30)$&$-0.19(27)$      &$0.002(283)$ &$-0.7(312.9)$&$0.5(349.6)$&$-0.5(103.8)$ 
&\\\hline
$\Xi^{0}_{c}\to n \eta^\prime $  &$0.00033(20)$&$0.96(57)$      & $0.0002(381)$&$-0.04(1700)$&$-1(78)$&$0.1(379.8)$  
&\\\hline
\hline
\end{tabular}
\end{table*}

\begin{table*}[htbp!]
\caption{The predicted values for branching ratios, polarization parameters ($\alpha,\beta,\gamma$) and CP violation
for different fits. Case I(II)[III] means the fit results for the real(complex)[new] form factors. 
}\label{tableccomplex}\begin{tabular}{|c|c|c|c|c|c|c|c|c|c|c|c}\hline\hline
\multirow{2}{*}{channel} &  \multicolumn{2}{c|}{ Case I}&  \multicolumn{4}{c|}{ Case II} 
& \multicolumn{1}{c|}{ Case III}\cr\cline{2-8}   & 
Br($\%$)& $\alpha$ & 
Br($\%$)& $\alpha$ & $\beta$ & $\gamma$ 
& CPV  
 \\\hline \hline
$\Lambda^{+}_{c}\to \Sigma^{0}  \pi^{+} $ &$1.248(46)$&$-0.472(15)$&$1.245(48)$ &$-0.472(15)$ & $-0.88(13)$&$-0.09(1.21)$
&\\\hline
$\Lambda^{+}_{c}\to \Lambda  \pi^{+} $ 
& $1.309(47)$&$-0.7536(60)$& $1.272(48)$&$-0.7551(60)$&$0.58(46)$&$0.30(89)$
& \\\hline
$\Lambda^{+}_{c}\to \Sigma^{+}  \pi^{0} $ &$1.262(42)$&$-0.470(15)$&$1.253(48)$ &$-0.471(15)$&$-0.88(13)$ &$-0.1(1.2)$ 
&\\\hline
$\Lambda^{+}_{c}\to p  K_{S}^{0} $ &$1.581(47)$& $0.39(14)$&$1.580(69)$ &$-0.05(38)$&$-0.8(1.6)$&$-0.6(1.9)$
&\\\hline
$\Lambda^{+}_{c}\to \Xi^{0}  K^{+} $ &$0.423(29)$&$0.957(18)$&$0.555(70)$ &$0.04(15)$&$0.37(88)$&$0.93(36)$
&\\\hline
$\Xi^{+}_{c}\to \Sigma^{+}  K_{S}^{0} $ &$0.81(21)$&$0.63(19)$&$0.9(2.0)$ &$0.81(35)$&$0.3(2.0)$&$0.49(97)$
&\\\hline
$\Xi^{+}_{c}\to \Xi^{0}  \pi^{+} $ &$0.875(77)$&$-0.900(38)$&$1.94(59)$ &$0.20(17)$&$0.4(1.2)$&$0.91(45)$
&\\\hline
$\Xi^{0}_{c}\to \Sigma^{0}  K_{S}^{0} $ &$0.054(16)$&$-0.41(29)$&$0.055(16)$ & $0.4(645.1)$ &$0.7(482.9)$&$0.6(526.4)$
&\\\hline
$\Xi^{0}_{c}\to \Lambda  K^0_S$  &$0.225(34)$&$0.88(18)$ &$0.324(60)$ & $-0.4(250.9)$&$0.9(109.4)$ &$-0.1(121.6)$
&\\\hline
$\Xi^{0}_{c}\to \Sigma^{+}  K^{-} $ &$0.195(39)$& $0.87(33)$&$0.181(40)$ &$0.9(326)$&$-0.07(1200)$&$-0.5(443.2)$
&\\\hline
$\Xi^{0}_{c}\to \Xi^{-}  \pi^{+} $ & $1.16(18)$&$-0.709(45)$& $1.18(19)$&$-0.640(51)$& $-0.76(10)$&$-0.08(89)$
& \\\hline
$\Xi^{0}_{c}\to \Xi^{0}  \pi^{0} $ &$0.152(48)$& $-0.45(13)$&$0.68(14)$&$-0.87(28)$& $0.39(53)$ &$-0.31(82)$
&\\\hline\hline
$\Lambda^{+}_{c}\to \Sigma^{0}  K^{+} $   &$0.0377(18)$&$-0.9960(42)$ &$0.0389(25)$ &$-0.58(16)$& $-0.2(1.1)$&$-0.79(34)$
&0.0015(18)\\\hline
$\Lambda^{+}_{c}\to \Lambda  K^{+} $  &$0.0639(29)$& $-0.547(44)$&$0.0645(31)$ &$-0.585(49)$&$-0.08(91)$&$0.806(99)$
&0.0011(13)\\\hline
$\Lambda^{+}_{c}\to \Sigma^{+} K^0_{S,L} $  &$0.0277(27)$& $-0.69(10)$&$0.042(13)$& $-0.81(46)$ &$-0.53(62)$&$0.26(71)$
&-0.0004(46)\\\hline
$\Lambda^{+}_{c}\to p  \pi^{0} $          &$0.0163(60)$&$-0.16(13)$&$0.0158(75)$ & $0.7(4.3)$&$-0.6(4.5)$&$0.4(1.2)$
& \\\hline
$\Lambda^{+}_{c}\to n  \pi^{+}$           &$0.0642(23)$&$0.539(43)$&$0.0766(88)$ & $0.10(17)$ &$0.5(1.6)$&$0.88(83)$
&-0.0052(39)\\\hline
$\Xi^{+}_{c}\to \Sigma^{0}  \pi^{+} $      & $0.3161(88)$& $-0.729(17)$&$0.254(13)$&$-0.902(31)$&$-0.35(34)$& $0.25(47)$
&\\\hline
$\Xi^{+}_{c}\to \Lambda  \pi^{+} $    &$0.0218(32)$&$-0.16(18)$&$0.095(10)$ &$0.43(31)$&$0.52(69)$&$-0.74(56)$ 
&-0.0005(32)\\\hline
$\Xi^{+}_{c}\to \Sigma^{+}  \pi^{0} $  &$0.288(19)$& $0.40(13)$&$0.263(94)$ &$0.27(56)$&$-0.7(1.3)$& $-0.6(1.4)$ 
&-0.00027(85)\\\hline
$\Xi^{+}_{c}\to p  K^0_{S,L} $          &$0.144(14)$&$-0.495(86)$&$0.155(24)$ &$-0.83(42)$&$-0.54(65)$&$-0.14(75)$ 
&-0.0002(27)\\\hline
$\Xi^{+}_{c}\to \Xi^{0}  K^{+} $         & $0.1340(55)$&$0.376(33)$&$0.162(33)$&$0.07(11)$&$0.3(1.2)$&$0.94(39)$ 
&-0.0052(39)\\\hline
$\Xi^{0}_{c}\to \Sigma^{0}  \pi^{0} $    &$0.00007(22)$ &$-0.5(2.3)$&$0.04(5.31)$&$-0.6(228.9)$&$0.8(151.0)$ &$-0.2(173.1)$ 
&-0.0009(42)\\\hline
$\Xi^{0}_{c}\to \Lambda  \pi^{0} $    &$0.0314(56)$&$0.71(12)$ &$0.03(5.47)$ &$-1(17)$&$-0.1(27.3)$&$-0.1(165.1)$ 
&0.0013(20)\\\hline
$\Xi^{0}_{c}\to \Sigma^{+}  \pi^{-} $     &$0.0123(26)$&$0.84(35)$&$0.01(33)$&$0.8(337.2)$&$-0.07(1200)$&$-0.6(412.1)$
&\\\hline
$\Xi^{0}_{c}\to p  K^{-} $                &$0.0154(40)$&$0.73(38)$&$0.01(1.50)$ &$0.7(349)$&$-0.06(1000)$&$-0.7(316.2)$ 
&\\\hline
$\Xi^{0}_{c}\to \Sigma^{-}  \pi^{+} $     &$0.0625(97)$&$-0.778(44)$&$0.071(16)$ &$-0.627(92)$&$-0.75(19)$&$-0.22(85)$ 
&\\\hline
$\Xi^{0}_{c}\to n  K^0_{S,L} $& $0.0201(50)$&$0.40(33)$&$0.02(81)$&$-0.4(507.4)$ &$0.7(345.2)$&$-0.6(213.4)$
&-0.0031(43)\\\hline
$\Xi^{0}_{c}\to \Xi^{-}  K^{+} $          &$0.0515(80)$&$-0.665(44)$&$0.0491(82)$ & $-0.642(71)$&$-0.767(54)$&$0.007(894)$
&\\\hline
$\Xi^{0}_{c}\to \Xi^{0}  K^0_{S,L} $  &$0.0088(20)$&$0.58(43)$&$0.01(83)$ &$-0.4(618.2)$&$0.9(346.5)$&$1(0)$ 
&-0.0051(66)\\\hline\hline
$\Lambda^{+}_{c}\to p  K^0_L $ 	&$1.689(49)$&$0.46(14)$	& $1.677(69)$ & $0.03(41)$&  $-0.8(1.7)$ &  $-0.6(2.0)$
&\\\hline
$\Lambda^{+}_{c}\to n  K^{+} $ 		&$0.001008(89)$&$-0.979(19)$&$0.00193(36)$ &$0.25(25)$&$0.5(1.4)$&$0.85(71)$
& \\\hline
$\Xi^{+}_{c}\to \Sigma^{0}  K^{+} $ & $0.01142(30)$&$-0.9961(13)$&$0.01006(53)$
&$-0.990(16)$&$0.12(13)$&$-0.07(25)$ 
&\\\hline
$\Xi^{+}_{c}\to \Lambda  K^{+} $&$0.00437(18)$&$0.620(30)$&$0.0047(10)$ &$0.28(14)$&$-0.86(62)$&$0.4(1.2)$ 
&\\\hline
$\Xi^{+}_{c}\to \Sigma^{+}  K^0_L $ & $1.01(23)$& $0.82(15)$&$1.1(2.1)$&$0.95(21)$&$0.2(2.2)$& $0.3(1.0)$ 
&\\\hline
$\Xi^{+}_{c}\to p  \pi^{0} $ 		&$0.00103(32)$& $0.10(26)$&$0.0014(27)$&$0.3(3.8)$&$0.2(1.9)$& $-0.9(1.5)$
&\\\hline
$\Xi^{+}_{c}\to n  \pi^{+} $ 		&$0.00612(39)$&$0.942(20)$&$0.0045(19)$ &$0.07(26)$&$0.7(1.3)$&$0.8(1.1)$
&\\\hline
$\Xi^{0}_{c}\to \Sigma^{0}  K_{L}^{0}$ &$0.069(18)$&$-0.14(29)$&$0.08(46)$ &$0.3(583.5)$&$0.8(319.9)$&$0.6(423.3)$ 
&\\\hline
$\Xi^{0}_{c}\to \Lambda  K^0_L$  &$0.211(31)$& $0.91(15)$&$0.31(15)$&$-0.4(207.9)$&$0.9(85.8)$& $-0.09(101.86)$
&\\\hline
$\Xi^{0}_{c}\to p  \pi^{-} $ 		&$0.00092(25)$& $0.71(38)$&$0.0009(1009)$ &$0.7(347.7)$&$-0.06(990)$& $-0.7(298.3)$
&\\\hline
$\Xi^{0}_{c}\to \Sigma^{-}  K^{+} $ &$0.00282(44)$&$-0.744(44)$&$0.00305(56)$&$-0.635(65)$&$-0.76(15)$&$-0.15(87)$ 
&\\\hline
$\Xi^{0}_{c}\to n  \pi^{0} $ 		&$0.00142(27)$&$0.998(18)$&$0.001(601)$&$-1(69)$&$0.2(87.3)$&$-0.09(525.77)$
&\\\hline
\hline
\end{tabular}
\end{table*}

\section{CPV analysis  under some scenarios }

In the above analysis, we present an analysis using the SU(3) invariant representation (IRA) method for anti-triplet charmed baryon two-body decays.
By incorporating the complex form factors, the nine IRA amplitudes can accurately reproduce the current experimental data.
The fitted phase of the form factor enables the study of CP violation (CPV) effects.
However, in previous work~\cite{Xing:2024nvg}, the complex form factor in the IRA method, with contributions from the penguin diagram, introduces 51 parameters, which cannot be determined with the current 37 experimental data.
 In our work, based on the analysis of the equivalence between TDA and IRA, we can predict the CP violation (CPV) under certain scenarios.



The above analysis shows that the IRA form factors with complex phases will introduce 35 parameters in our fit. 
To reduce the number of parameters, we can determine the strong phase from the perspective of topological diagrams, under the assumption that the imaginary part of the form factor in Eq.~\ref{tdaff}, corresponding to each topological diagram in Fig.\ref{fig1}, is equal.
Since the strong phase originates from the potential contributions of intermediate on-shell states in the decay process~\cite{PDG}, this assumption appears to be reasonable.
In this analysis, we further assume $\mathcal{I}m(f_{T_3})=0$ due to the global phase.


Since the new form factor may introduce a new weak phase, the CPV of the anti-triplet charmed baryon two-body decays can be studied under the assumption that the new weak phase is solely caused by the new form factors.
To investigate CP violation (CPV), the potential origins of the new form factor are analyzed. The form factor $\textbf{f/g}_b$ arises from the difference between topological diagrams $T_1$ and $T_2$, which have distinct color structures shown in Fig.~\ref{fig1}. The most likely source is the non-factorizable QCD contribution.
The SU(3) symmetry breaking effect could also be the source, as the new form factor reveals the discrepancy between the SU(3) analysis methods: TDA and IRA.
Particularly, when long-distance interactions influence the interaction vertex in Fig.~\ref{fig1}, the flavor symmetry breaking effect is inevitably involved.
Furthermore, final state rescattering (FSR) is anticipated to play a significant role in the new form factor, as these effects are not included in the IRA amplitude. 
Recent studies indicate that final state rescattering effects play a significant role in determining the magnitude of CP asymmetries in charmed hadron decay processes~\cite{He:2024pxh, Jia:2024pyb}.
 Since it can contribute to topological diagrams involving singly Cabibbo-suppressed processes, final state rescattering is likely a potential source of the new form factor.
In addition, the penguin operator can also contribute to the new form factor $\textbf{f}_b$. Under flavor symmetry, the difference in the Lorentz structure at the Hamiltonian vertex can be ignored. Consequently, the penguin operator has the same Hamiltonian matrix structure as $O_{1,2}$
  in TDA. In IRA, the penguin operator only contributes to $H_3$, which we neglect in our analysis.
Consequently, the new form factor $\textbf{f}_b$, arising from the conflicting parts of IRA and TDA, must be accounted for. 
Since multiple potential sources could lead to new form factors, our approach is not restricted to a single CP violation (CPV) origin. We will examine all possible sources and incorporate the new form factors into our fit, performing a numerical analysis to assess its impact.


Based on the potential origins of these new form factors, their contributions can be estimated by referring to previous studies. The contribution of the penguin operator is found to be less than 10\%~\cite{Wang:2024ztg, Xing:2024nvg}.
Although there are many indications suggesting SU(3) symmetry-breaking effects in charmed baryon decays, the previous global fit still shows that the symmetry-breaking effect is negligible within its error margins~\cite{He:2021qnc}.
Regarding the FSR, recent studies suggest that the strong phase in the parity-conserving form factor related to FSR is nearly zero~\cite{He:2024pxh}.
Therefore, in our work, we can neglect the imaginary component of the parity-conserving new form factor as ${\mathcal I}m(\textbf{g}_{b/c})=0$, since $\textbf{g}_{b/c}$  appears in conjunction with other TDA form factors. 
However, the new form factor $\textbf{f}_b$ exclusively contributes to the IRA form factor $f_6^b$, suggesting that its contribution cannot be ignored.

According to the above analysis, the new form factor can be expressed as 
\begin{eqnarray}
 \textbf{f}_b=\textbf{f}^S_b+\textbf{f}^P_b e^{i\phi^P}\;,
\end{eqnarray}
where $\textbf{f}^S_b$ is derived from SU(3) symmetry breaking and non-factorizable QCD contributions, which do not produce a weak phase.
 The term $\textbf{f}^P_b e^{i\phi^P}$ represents the contribution from the penguin operator and final state radiation (FSR) effects, with the weak phase $\phi^P=-1.147\pm 0.026$. 
 The form factors in our global analysis are expressed as:
\begin{eqnarray}
&&A^q_{6,15}=e^{i\phi_1}\bigg(\mathcal{R}e(A^q_{6,15})+\mathcal{I}m(A^q_{6,15})\bigg)
,\notag\\
&&f^b_{6}=\frac{3}{2}\bigg(e^{i\phi_1}(\mathcal{R}e(\textbf{f}^S_b)+\mathcal{I}m(\textbf{f}^S_b))\notag\\
&&\quad\quad+e^{i\phi^P}(\mathcal{R}e(\textbf{f}^P_b)+\mathcal{I}m(\textbf{f}^P_b))\bigg),\notag\\
&&f^d_6=e^{i\phi_1}(\mathcal{R}e(f^d_6)+\mathcal{I}m(f^d_{6}))\notag\\
&&\quad\quad-e^{i\phi^P}\frac{3}{2}(\mathcal{R}e(\textbf{f}^P_b)+\mathcal{I}m(\textbf{f}^P_b)),\notag\\
&&f^e_{15}=e^{i\phi_1}(\mathcal{R}e(f^e_{15})+\mathcal{I}m(f^e_{15}))\notag\\
&&\quad\quad-e^{i\phi^P}\frac{3}{2}(\mathcal{R}e(\textbf{f}^P_b)+\mathcal{I}m(\textbf{f}^P_b)),\;A=f,g,
\end{eqnarray}
where $q=a,b,c,d,e$. 
The weak phase $\phi_1$ arises from the current-current operator as $\phi_1=\arg(V_{cq}^*V_{uq'})\approx 0,-\pi$. On can find that the penguin operator only contribute to the Cabibbo-suppressed processes. The SU(3) symmetry requires the  form factor corresponding to different processes to be equal~\cite{Geng:2023pkr,Zhong:2024qqs}, one has the chance to determine the form factor $\textbf{f}^S_b$ by the Cabibbo-allowed and doubly Cabibbo-suppressed processes. Based on the symmetry breaking form factor we derived, the penguin contribution form factor $\textbf{f}^P_b$ can be further determined by the Cabibbo-suppressed processes.

By introducing the weak phase of penguin operator $\phi^T$, one can derive the CPV if the global fit gives a nonzero value of $\textbf{f}^P_b$.
In our analysis, we obtains the fit results with  $\chi^2/d.o.f=1.004$, indicating that our fit is reasonable and the form factors effectively explain the experimental data.

With the determined penguin operator contribution $\mathcal{R}e(\textbf{f}^P_b)=0.000051(40)$ and $\mathcal{I}m(\textbf{f}^P_b)=-0.000049(80)$, one can predict the CP violation ($A_{CP}$) for the Cabibbo-suppressed processes.  Considering the ratio of the form factor from current-current and penguin operator $\frac{\langle O_{1,2}\rangle}{\langle O_{3-6}\rangle}\sim O(10^{-2})$, our study implies that the contribution of the QCD penguin may be more substantial than conventionally assumed, necessitating further theoretical investigation in the future.

Based on the fitted results, we predict the values of $A_{CP}$ for the Cabibbo-suppressed processes and find that almost all the predicted $A_{CP}$ values are approximately zero within the errors, except for the following processes
 \begin{eqnarray}\label{acp}
&&
A_{CP}^{\Lambda_c^+\to n\pi^+}=-0.0052(39),\notag\\
&&A_{CP}^{\Xi_c^+\to \Xi^0 K^+}=-0.0052(39).
\end{eqnarray}
We strongly recommend the experimental measurement of the  processes: $\Lambda_c^+\to n \pi^+$ since their  branching ratio has been  measured.
Surprisingly,
one can observe that the CPV we predicted for $\Lambda_c^+\to n\pi^+$ and $\Xi_c^+\to \Xi^0 K^+$ channels are equal.
The identical CPV effects in the $\Lambda_c^+\to n\pi^+$ and $\Xi_c^+\to \Xi^0 K^+$ decay channels arise from their shared SU(3) decay amplitudes, expressed as $-\sin\theta (c_6-c_{15}+d_6+d_{15})$.
Moreover, this analysis assumes that the new weak phase is introduced exclusively through the new form factor $\textbf{f}b$, thereby restricting CPV contributions to specific IRA form factors $f^b_6$, $f^d_6$, and $f^e{15}$. Consequently, the amplitudes for these decay channels rely on the same combination, $d_6+e_{15}$, which results in identical CPV effects.
The CPV for other channels (case III) is summarized in Table.~\ref{tableeta}  and Table.~\ref{tableccomplex}.

\section{Conclusion}

In this study, we performed a global analysis of two-body decays of anti-triplet charmed baryons. Utilizing 35 experimental data points, we successfully determined 18 real IRA form factors, as shown in Table~\ref{table2} (Case I).
The branching ratios and polarization parameters (Case I) for these processes are presented in Table~\ref{tableeta}  and Table~\ref{tableccomplex}, respectively. Notably, the polarization parameters of $\Lambda_c^+ \to \Xi^0 K^+$ and $\Xi_c^0 \to \Xi^0 \pi^0$ show discrepancies when compared with our results, indicating the necessity of considering complex form factors.

Fortunately, by incorporating the complex 18 IRA form factors, the 35 parameters can be determined using a total of 37 experimental measurements. The fitted form factors and predictions (Case II) are provided in Table.~\ref{table2} , Table.~\ref{tableeta}, and Table.~\ref{tableccomplex}, respectively. 
Although the low degrees of freedom result in large uncertainties, the $\chi^2/\mathrm{d.o.f}=1.36$ indicates that SU(3) symmetry is well-preserved.

Based on the form factors determined in the global fit (Case I), we analyze the equivalence between the SU(3) IRA and TDA methods.
Our findings suggest that the results derived from the IRA method are largely consistent with the assumptions of the TDA method, as depicted in the topological diagram in Fig.~\ref{fig1}.
However, some form factors exhibit deviations from the predicted values based on the topological diagrams. To quantify these deviations, we introduce the form factors $\textbf{f}_b$ and $\textbf{g}_{b/c}$ to represent the new effects.

Leveraging the newly introduced form factors, we aim to investigate the CPV effect under specific scenarios. 
Assuming that the imaginary part of the form factor in Eq.~\ref{tdaff} is equal for each topological diagram in Fig.\ref{fig1}, the number of parameters can be reduced to 31.
Furthermore, assuming that the new weak phase arises solely from the new form factor, 
the CPV in anti-triplet charmed baryon two-body decays can be predicted. The fitted form factors and the predicted CPV values (case III) are presented in Table.\ref{table2}, Table.\ref{tableeta}, and Table.\ref{tableccomplex}, respectively.
Based on these assumptions, our analysis predicts non-zero $A_{CP}$ in Eq.~\ref{acp}. Given its associated error, observable CP violation can be detected at the level of $O(10^{-3})$. We strongly recommend measuring these processes.


It is important to note that the most ideal condition for measuring CP violation is the simultaneous production of charmed baryon 
and anti-baryon pairs. 
The BESIII Collaborations meet these conditions, with extensive data on $\Lambda_c \bar \Lambda_c$  and $\Xi_c \bar \Xi_c$ pairs. 
For $\Lambda_c^+\to n \pi^+$, the branching ratio has already been measured by the BESIII Collaborations. 
Therefore, this facility holds the potential to observe CP violation in charmed baryon decays for the first time.

\section*{Acknowledgements}

We thank Prof. Xiao-Gang He, Prof. Wei Wang and Prof. YuJi Shi for useful discussion.
The work of Jin Sun is supported by IBS under the project code, IBS-R018-D1. 
The work of Ruilin Zhu is supported by NSFC under grant No. 12322503 and No. 12075124, and by Natural Science Foundation of Jiangsu under Grant No. BK20211267.
The work of Zhi-Peng Xing is supported by NSFC under grant No.12375088 and No. 12335003.

\bibliographystyle{JHEP}
\bibliography{ref}

\end{document}